\documentclass[letterpaper,10pt,conference]{ieeeconf}
\IEEEoverridecommandlockouts
\usepackage[T1]{fontenc}
\usepackage{amsmath,amssymb,amsfonts,mathrsfs,booktabs,graphicx,cite,url}
\let\labelindent\relax
\usepackage{microtype,tikz,xcolor,enumitem,etoolbox}
\usetikzlibrary{arrows.meta,positioning,fit,backgrounds}

\newcommand{\Sym}{\operatorname{Sym}}
\newcommand{\diag}{\operatorname{diag}}
\newcommand{\col}{\operatorname{col}}
\newcommand{\tr}{\operatorname{tr}}
\newcommand{\R}{\mathbb R}
\newtheorem{theorem}{Theorem}
\newtheorem{lemma}{Lemma}
\newtheorem{corollary}{Corollary}
\newtheorem{proposition}{Proposition}
\newtheorem{assumption}{Assumption}

\newenvironment{proof}{\par\noindent\textit{Proof:} }{\hfill$\square$\par}
\setlist{nosep,leftmargin=*}
\AtBeginEnvironment{theorem}{\setlength{\topsep}{2pt}\setlength{\partopsep}{0pt}}
\AtBeginEnvironment{lemma}{\setlength{\topsep}{2pt}\setlength{\partopsep}{0pt}}
\AtBeginEnvironment{corollary}{\setlength{\topsep}{2pt}\setlength{\partopsep}{0pt}}
\AtBeginEnvironment{proposition}{\setlength{\topsep}{2pt}\setlength{\partopsep}{0pt}}
\AtBeginEnvironment{assumption}{\setlength{\topsep}{2pt}\setlength{\partopsep}{0pt}}
\AtBeginEnvironment{proof}{\setlength{\topsep}{2pt}}

\AtBeginEnvironment{thebibliography}{\footnotesize\setlength{\itemsep}{0pt}\setlength{\parsep}{0pt}}

\makeatletter
\def\section{\@startsection{section}{1}{\z@}{0.8ex plus .4ex minus .3ex}{0.3ex plus .3ex}{\normalfont\normalsize\centering\scshape}}
\def\subsection{\@startsection{subsection}{2}{\z@}{0.8ex plus .4ex minus .3ex}{0.3ex plus .3ex}{\normalfont\normalsize\itshape}}
\makeatother
\allowdisplaybreaks[2]

\title{Local Certificate Repair for Asynchronous Controller Updates in Recommender Systems}
\author{Simone Mariano and Paolo Frasca\thanks{The authors are with Univ. Grenoble Alpes, CNRS, Inria, Grenoble INP, GIPSA-lab, 38000 Grenoble, France. This work has been supported in part by the ANR project BEAR (ANR-22-CE38-0017-01).}}
\begin{document}
\maketitle
\begin{abstract}
We study how to adapt an existing distributed recommendation controller when local dynamics or objective weights change. Starting from a quadratic stability and cost certificate, we construct positive semidefinite corrections with finite spatial support and an explicit residual. A joint admission test allocates the available dissipation margin to these residuals and to interactions among changes and corrections, including pending updates. The test covers complete declared parameter ranges, allowing different operators to progress independently. For a fixed compatible batch, the corrections define a common storage that guarantees exponential stability and bounds the cost accumulated throughout deployment. When the admitted collection evolves, a reservation protocol preserves the certificate and a rate bound on storage increases ensures exponential convergence. Regional linear-quadratic regulator (LQR) design proposes new gain rows for the operators’ own users, and a separate residual-based calculation can certify that the final gain lowers the expected cost. A recommendation-network example illustrates the tradeoff between repair depth, data access, and the benefits of deploying an update.
\end{abstract}


\section{Introduction}

Recommendation systems close a feedback loop: the content shown to users influences their opinions and interactions, which in turn affect future recommendations~\cite{DePasquale2026Control, Mariano2026Optimal}. On large platforms, a multitude of users interact with one another and with the recommender, which acts as a controller, forming a large-scale networked system. Requiring each recommendation to depend on the instantaneous states of all other users would be impractical and unreliable at this scale. The recommendation system must therefore be distributed, in the sense that the recommendation to each user depends only on the states of a prescribed subset of users. In~\cite{Mariano2026Structured}, such a distributed controller is a linear gain obtained through a linear-quadratic regulator (LQR) formulation that balances opinion alignment, polarization, and recommendation effort, and it comes with a certificate of stability and cost for the nominal closed loop.

In this context, a controller designed for one operating condition must be adapted when that condition changes, and on social platforms such changes are often local. Preferences drift over time~\cite{Koren2009}, and around major real-world events, studies report changes in communication patterns and a concentration of attention on particular users~\cite{Lin2014Attention}, bursts of connection creation and deletion~\cite{Myers2014}, and changes in online hostility that depend on users' network environments~\cite{Rasmussen2023}. These observations motivate revising the network model and the weights used in the design of recommendation systems.

Concretely, the interaction parameters of a group of users may change, or the platform may change the weights assigned to their recommendations. In these scenarios, redesigning the whole distributed controller would require global knowledge and computation. Instead, each operator, namely a party that is responsible for the recommendations to one group of users and sees only local data, proposes new gain rows for its group and leaves the other rows fixed. Because operators act at different times, the network passes through mixed configurations that a certificate for the final gain alone does not cover. The question is therefore how to preserve stability and a cost bound throughout these asynchronous updates, using only local data and the inherited certificate. This question hides two obstructions. First, even when a change is confined to one group of users, its effect on the certificate is not. The closed loop couples users along the network, and the inherited storage may couple them further through cross-user terms, so an exact repair requires solving a Lyapunov-type equation whose solution is dense and therefore global. Second, when several operators change their groups independently, all repairs draw on the same dissipation margin, so checks that pass one by one can fail jointly. A local change in a coupled network is thus not simply a local re-certification problem.

Existing results avoid these obstructions by assumption rather than resolving them. Retrofit control~\cite{Ishizaki2021} and distributed dissipativity synthesis~\cite{Agarwal2021} make certificates compositional by design, through a structural constraint on the local controller or through common local storage matrices that exclude cross-user terms. They therefore do not apply to a gain and certificate inherited from a design that imposes neither structure. Localized stability conditions~\cite{Motee2022}, low-rank matrix-equation updates~\cite{Kressner2019}, and sparse Lyapunov approximations~\cite{Haber2016} localize or cheapen the certificate of a given system but, to our knowledge, do not account for the margin shared by several concurrent changes. Localized LQR~\cite{Shin2023}, distributed policy gradient~\cite{Yan2024}, and asynchronous parallel LQR learning~\cite{Zhao2025} compute localized policies or tolerate delays in computing them, but they do not address whether the mixed configurations reached during deployment remain certified. What is missing is a way to repair a coupled certificate from local data and to account for all changes in progress at once, and this is what we provide. 

We remove these obstructions starting from the gain and certificate inherited from~\cite{Mariano2026Structured}. Each change is declared through an admissible parameter range, supplied by a revised model, a variation bound, or a planned intervention. The range fixes what must be certified but not when the change starts or how it progresses, so operators need not track one another's parameter progress. Planned changes may wait for admission, whereas an external change, which no operator controls, must already be covered by a certificate, and model identification and change detection are upstream tasks. The dense exact repair is replaced by a finite Gramian, which computes from local data a positive semidefinite correction with finite support and an explicit residual. The shared margin is managed by a joint admission test, which charges residuals and the interactions between corrections and changes against it and reserves margin before installation, so changes admitted separately cannot conflict. The mixed configurations are covered by an assembly theorem, which gives a common certificate for every configuration of an admitted batch and a bound on the cost of deploying it, and by a reservation protocol with an average-rate condition on storage increases, which extends stability to collections that keep evolving. Finally, value approximations with residual bounds can certify that the final configuration lowers the expected cost.

We first formalize the model and the update problem (Section~\ref{sec:model}). We then repair the certificate locally and test changes jointly (Section~\ref{sec:repair}), deploy the repairs asynchronously and verify the resulting cost improvement (Section~\ref{sec:protocol}), and illustrate the procedure on a recommendation network (Section~\ref{sec:example}). Section~\ref{sec:disc} provides the conclusion and perspectives.

\emph{Notation.} $B\succeq D$ means $B-D$ is positive semidefinite. $B_+$ replaces the negative eigenvalues of a symmetric matrix $B$ by zero, so $B\preceq B_+$ and $B_+\succeq0$. We use the Euclidean and induced norms, $\tr$ for trace, $\Sym(B)=B+B^\top$, and $\diag$ for diagonal or block-diagonal assembly. $\lambda_{\min}$ and $\lambda_{\max}$ denote the smallest and largest eigenvalues, respectively, and $\mathbb N$ includes zero.
Subscripts $B_{\mathcal U\mathcal W}$ select the rows and columns associated with users in $\mathcal U$ and $\mathcal W$, including all their topics, and a colon selects all users. A matrix has square support $\mathcal U$ when its blocks outside $\mathcal U\times\mathcal U$ vanish. The diagonal projector $I_{\mathcal U}$ keeps the states of users in $\mathcal U$ and sets all other components to zero.

\section{Model, information architecture, and local updates}\label{sec:model}
This section sets up the problem. We first recall the opinion model and the objective, then describe how the recommendations are split among operators that have limited access to information, and recall the certificate of the controller to be updated. Finally, we describe how an operator proposes a local update and state the certification problem addressed in the rest of the paper.

\subsection{Opinion dynamics and cost function}
Consider $n$ users holding opinions on $m$ topics. Their interactions are described by the directed weighted graph $\mathscr G_{\mathrm{soc}}=(\mathcal V,\mathcal E_{\mathrm{soc}})$, where $\mathcal V=\{1,\ldots,n\}$ and $a_{ij}\ge0$, positive exactly when $(j,i)\in\mathcal E_{\mathrm{soc}}$, measures the influence of user $j$ on user $i$. Let $\Gamma=[a_{ij}]$ be its adjacency matrix and $L=\diag(\Gamma\mathbf 1_n)-\Gamma$ its Laplacian. For each user $i$, let $z_i,w_i\in\R^m$ denote the opinion and recommendation vectors, stacked as $z=\col(z_i)$ and $w=\col(w_i)$ in $\R^p$, with $p=nm$. Following~\cite{Mariano2026Structured}, the opinion dynamics and feedback law are
\begin{equation}
\dot z=Az+(A_a\otimes I_m)z_a+w,\qquad w=-Fz,\label{eq:model}
\end{equation}
\begin{equation}
A=I_n\otimes(C-I_m)-(L+A_a)\otimes I_m-I_p.\label{eq:drift}
\end{equation}
Here $C\in\R^{m\times m}$ describes interactions between topics, $A_a=\diag(a_i)$ with $a_i>0$ collects the anchoring strengths, and $z_a$ collects anchor opinions. Recommendations act through the difference $w-z$, whose contribution $-z$ is included in $A$. Throughout, $z_a=0$, so the closed-loop dynamics are $\dot z=Mz$ with $M=A-F$, and every configuration shares the origin as an equilibrium. With zero anchor opinions, the stage cost of~\cite{Mariano2026Structured} is
\begin{align}
\ell(z,w)={}&-z^\top W_{\mathrm{EN}}w+z^\top(W_{\mathrm P}+W_{\mathrm D})z
\nonumber\\
&+w^\top W_{\mathrm{EX}}w+(z-w)^\top W_{\mathrm C}(z-w)
\nonumber\\
&+\alpha_F w^\top(L_F\otimes I_m)w.\label{eq:stage-components}
\end{align}

The first term rewards opinion--recommendation alignment as a proxy for engagement. The matrices $W_{\mathrm P}$ and $W_{\mathrm D}$ weight penalties on polarization and on deviation from the zero uncontrolled equilibrium, $W_{\mathrm{EX}}$ penalizes recommendation effort, and $W_{\mathrm C}$ penalizes recommendation--opinion mismatch. These matrices are diagonal, with $W_{\mathrm{EN}},W_{\mathrm P},W_{\mathrm D}\succeq0$ and $W_{\mathrm{EX}},W_{\mathrm C}\succ0$. The last term, with weight $\alpha_F\ge0$, promotes exposure coherence through the Laplacian $L_F$ of an undirected exposure graph. Following~\cite{Mariano2026Structured}, let $\mathcal F_1,\ldots,\mathcal F_r$ be a partition of $\mathcal V$ into exposure groups, each inducing a connected subgraph of the symmetrized social graph. The exposure graph $\mathscr G_{\mathrm{exp}}=(\mathcal V,\mathcal E_{\mathrm{exp}})$ retains the social links within each group, $\mathcal E_{\mathrm{exp}}=\bigcup_{\ell=1}^{r}\{\{i,j\}:i,j\in\mathcal F_\ell,\ i\ne j,\ a_{ij}+a_{ji}>0\}$, with edge weights $(a_{ij}+a_{ji})/2$. Denoting its adjacency matrix by $\Gamma_{\mathrm{exp}}$, we set $L_F=\diag(\Gamma_{\mathrm{exp}}\mathbf 1_n)-\Gamma_{\mathrm{exp}}$.

Expanding~\eqref{eq:stage-components} gives
\begin{align*}
Q&=W_{\mathrm D}+W_{\mathrm P}+W_{\mathrm C},\qquad
N=-\tfrac12 W_{\mathrm{EN}}-W_{\mathrm C},\\
R&=W_{\mathrm{EX}}+W_{\mathrm C}+\alpha_F(L_F\otimes I_m),
\end{align*}
so that $\ell(z,w)=z^\top Qz+2z^\top Nw+w^\top Rw$. The analysis below uses only this quadratic form: it admits general weights that satisfy Assumption~\ref{as:convex} and the support conditions on local changes introduced in Section~\ref{sec:envelope}. Under $w=-Fz$, the stage cost is $z^\top W(F;Q,N,R)z$, where
\begin{equation}
W(F;Q,N,R)=Q-NF-F^\top N^\top+F^\top RF,\label{eq:cost}
\end{equation}
or $W(F)$ when the weights are understood. If $M$ is Hurwitz, the cost from $z(0)=z_0$ is $z_0^\top\Pi_Fz_0$, with \emph{value matrix} $\Pi_F=\int_0^\infty e^{M^\top t}W(F)e^{Mt}\,dt$.
\begin{assumption}\label{as:convex}
Every objective $(Q,N,R)$ considered, before, during and after the changes, satisfies $Q=Q^\top$, $R=R^\top\succ0$ and $Q-NR^{-1}N^\top\succeq qI$ for a common $q>0$.
\end{assumption}
Completing the square in $w$ shows that $W(F)\succeq qI$ for every gain $F$, stabilizing or not.

\subsection{Operators, ownership, and information}
The controller to be updated is operated by several parties, which own different recommendations and have access to different information. Partition the users into nonempty disjoint slots $\{\mathcal B_b:b\in\mathcal B\}$, where $\mathcal B$ is the set of operators and $\mathcal V=\biguplus_{b\in\mathcal B}\mathcal B_b$. Operator $b$ controls the recommendations to the users in $\mathcal B_b$ and may therefore modify only the block rows $F_{\mathcal B_b,:}$.

Two further sets, with different roles, are attached to each operator. The feedback set $\mathcal H_b$ contains the users whose states may enter its recommendations, while the read set $\mathcal S_b$ contains the users whose data are acquired for redesign and certification. They satisfy $\mathcal B_b\subseteq\mathcal H_b\subseteq\mathcal S_b\subseteq\mathcal V$. Every implemented gain, inherited or updated, uses only the states of the users in $\mathcal H_b$ for the users in $\mathcal B_b$, that is, the block $F_{ij}$ vanishes whenever $i\in\mathcal B_b$ and $j\notin\mathcal H_b$. Thus $\mathcal H_b$ restricts the controller, whereas $\mathcal S_b$ governs only data access.

Operators communicate along an undirected operator graph $\mathscr G_{\mathrm{op}}=(\mathcal B,\mathcal E_{\mathrm{op}})$. Two operators $b\ne d$ are adjacent when a social link joins their slots, that is, $a_{ij}+a_{ji}>0$ for some $i\in\mathcal B_b$ and $j\in\mathcal B_d$, or when the feedback set of one contains users of the other, that is, $\mathcal H_b\cap\mathcal B_d\ne\varnothing$ or $\mathcal H_d\cap\mathcal B_b\ne\varnothing$. Adjacent operators exchange data directly, and nonadjacent operators use paths in this graph. The owners of all required data and resources must be reachable.

For the inherited controller $w=-F_0z$, designed for nominal data, the feedback constraint requires $(F_0)_{\mathcal B_b,j}=0$ whenever $j\notin\mathcal H_b$. The set $\mathcal H_b$ may include further users to allow new dependencies during redesign.

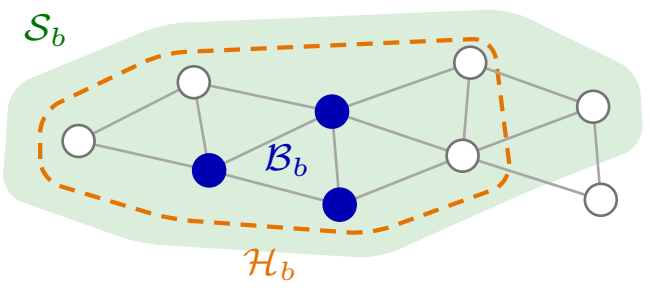
\begin{figure}[!t]
\centering
\begin{tikzpicture}[x=0.40in,y=0.255in,font=\scriptsize,
  u/.style={circle,draw=black!55,fill=white,inner sep=0pt,minimum size=6pt,line width=0.5pt},
  ob/.style={circle,fill=blue!70!black,inner sep=0pt,minimum size=6.5pt}]
  \coordinate (n1) at (0.0,0.75); \coordinate (n2) at (0.75,1.35); \coordinate (n3) at (0.85,0.45);
  \coordinate (n4) at (1.65,1.05); \coordinate (n5) at (1.70,0.10); \coordinate (n6) at (2.50,0.60);
  \coordinate (n7) at (2.55,1.55); \coordinate (n8) at (3.35,1.10); \coordinate (n9) at (3.40,0.15);
  \fill[green!50!black!14,rounded corners=4pt] (-0.50,0.20) -- (-0.45,1.30) -- (0.55,1.95) -- (2.80,2.10) -- (3.65,1.45) -- (3.68,0.70) -- (2.95,0.25) -- (1.95,-0.45) -- (0.40,-0.30) -- cycle;
  \draw[orange!90!black,densely dashed,line width=0.8pt,rounded corners=4pt] (-0.25,0.35) -- (-0.25,1.05) -- (0.55,1.65) -- (2.70,1.80) -- (2.82,0.32) -- (1.90,-0.20) -- (0.60,-0.05) -- cycle;
  \foreach \a/\b in {1/2,1/3,2/3,2/4,3/4,3/5,4/5,4/6,4/7,5/6,6/7,6/8,6/9,7/8,8/9}
    \draw[black!35,line width=0.5pt] (n\a) -- (n\b);
  \foreach \i in {1,2,6,7,8,9}{\node[u] at (n\i) {};}
  \foreach \i in {3,4,5}{\node[ob] at (n\i) {};}
  \node[blue!70!black] at (1.36,0.55) {$\mathcal B_b$};
  \node[orange!90!black,anchor=north] at (1.25,-0.16) {$\mathcal H_b$};
  \node[green!45!black,anchor=south west] at (-0.48,1.53) {$\mathcal S_b$};
\end{tikzpicture}
\caption{Ownership and information sets of operator $b$. The blue users form its slot $\mathcal B_b$, the orange dashed feedback set $\mathcal H_b$ contains the slot and its immediate neighbors, and the green shaded read set $\mathcal S_b$ contains the users whose data are read for redesign and certification.}
\label{fig:slot}
\end{figure}

The read set must contain the users indexing the rows and columns of every numerical block used for regional synthesis, defect bounds, repair, and interaction charges, including data supplied by other operators. It may grow up to $\mathcal V$ without changing $\mathcal H_b$ or ensuring admission. Figure~\ref{fig:slot} illustrates ownership, feedback access, and a possible read region.

\subsection{The inherited controller and its certificate}
Updates start from an existing controller together with a certificate of its performance. Let $(A_0,Q_0,N_0,R_0)$ be the nominal data for which $F_0$ was designed, and set $M_0=A_0-F_0$ and $W_0=W(F_0;Q_0,N_0,R_0)$. The pair $(F_0,P_0)$ may be obtained from~\cite{Mariano2026Structured}, while the subsequent analysis accepts any pair satisfying the following assumption.
\begin{assumption}\label{as:ref}
There are $P_0\succ0$ and a diagonal margin $\Sigma_0\succ0$ such that
\begin{equation}
\Sym(P_0M_0)+W_0\preceq-\Sigma_0.
\label{eq:anchor}
\end{equation}
\end{assumption}
Under Assumption~\ref{as:convex}, inequality~\eqref{eq:anchor} makes $M_0$ Hurwitz and gives $\Pi_{F_0}\preceq P_0$~\cite{Boyd1994}. The storage $V_0(z)=z^\top P_0z$ satisfies $\dot V_0+z^\top W_0z\le-z^\top\Sigma_0z$ along nominal trajectories. Thus $\Sigma_0$ assigns a dissipation margin to each user and topic, while $P_0$ may couple users.

\subsection{Proposals, declared changes, and the certification problem}
The certificate of Assumption~\ref{as:ref} holds only for the nominal data. When the model or objective changes, operator $b$ seeks to adapt the recommendations to the users in $\mathcal B_b$. Using the data available within $\mathcal S_b$, it proposes new gain rows subject to the feedback restriction of $\mathcal H_b$.

One option is to solve a regional LQR problem for the target data $(A,Q,N,R)$, the nominal data revised by its own change. Let $\widetilde F$ equal $F_0$ with block rows $\mathcal B_b$ set to zero, so that the other users keep their nominal policies, and let $u\in\R^{m|\mathcal B_b|}$ denote the recommendations of operator $b$. Restricting the state to the users in $\mathcal H_b$ and neglecting the contribution of the states of users outside this set gives the regional model
\begin{gather*}
\dot z_{\mathcal H_b}=A_b^{\mathrm{reg}}z_{\mathcal H_b}+B_b^{\mathrm{reg}}u,\\
A_b^{\mathrm{reg}}=(A-\widetilde F)_{\mathcal H_b\mathcal H_b},\qquad B_b^{\mathrm{reg}}=I_{\mathcal H_b\mathcal B_b}.
\end{gather*}
Here $I_{\mathcal H_b\mathcal B_b}$ is the submatrix of the identity with rows indexed by $\mathcal H_b$ and columns indexed by $\mathcal B_b$. Substituting the same input policies into the objective gives the regional cost matrices
\begin{gather}
Q_b^{\mathrm{reg}}=W(\widetilde F)_{\mathcal H_b\mathcal H_b},\qquad N_b^{\mathrm{reg}}=(N-\widetilde F^\top R)_{\mathcal H_b\mathcal B_b},\nonumber\\
R_b^{\mathrm{reg}}=R_{\mathcal B_b\mathcal B_b},
\label{eq:regional-cost}
\end{gather}
where $W(\widetilde F)$ is evaluated with the target weights $(Q,N,R)$. Since recommendations to users outside $\mathcal H_b$ may depend on states in $\mathcal H_b$, computing~\eqref{eq:regional-cost} may require gains and weights of users outside $\mathcal H_b$, which operator $b$ obtains through its read set.

If the corresponding continuous-time algebraic Riccati equation has a stabilizing solution $\Pi_b^{\mathrm{reg}}$, the optimal regional feedback is $u=-K_bz_{\mathcal H_b}$ with $K_b=(R_b^{\mathrm{reg}})^{-1}\big((B_b^{\mathrm{reg}})^\top\Pi_b^{\mathrm{reg}}+(N_b^{\mathrm{reg}})^\top\big)$. Extending $K_b$ by zero columns outside $\mathcal H_b$ gives $\bar K_b$. Operator $b$ then selects a step $\alpha_b\in[0,1]$ and proposes the gain change
\begin{equation*}
(\Delta F_b)_{\mathcal B_b,:}=\alpha_b\big(\bar K_b-(F_0)_{\mathcal B_b,:}\big),\qquad (\Delta F_b)_{\mathcal V\setminus\mathcal B_b,:}=0.
\end{equation*}
The choice $\alpha_b=1$ replaces the owned rows, while smaller steps blend the two policies. Deployment timing is independent of $\alpha_b$, and certification accepts any proposal satisfying the ownership, feedback, and data-access requirements.

A \emph{record} of operator $b$ collects a proposed change and the data needed to certify it. It declares a parameter set $\Theta_b$ containing all permitted intermediate configurations of its model, weights, and gain. The parameter $\theta_b\in\Theta_b$ may have several components, so these quantities can progress independently. The associated closed-loop increment $\Delta M_b(\theta_b)$ combines the model and feedback changes and has nonzero blocks only in $\mathcal B_b\times\mathcal D_b$, where $\mathcal D_b\subseteq\mathcal V$ contains all users whose states may enter the changed rows. Any addition or removal of social links must be included in this declared support.

A record passes through three stages, as formalized in Section~\ref{sec:protocol}: it is \emph{admitted} when certification resources are reserved for it, \emph{installed} when its certificate correction is published, and \emph{applied} when its physical change is allowed to progress. Each operator holds at most one admitted record, and $\mathcal R$, $\mathcal J$ and $\mathcal A$ denote the sets of operators whose records are admitted, installed and applied, respectively, with $\mathcal A\subseteq\mathcal J\subseteq\mathcal R$. Only applied changes act on the plant. Since the objective can couple recommendations of different operators, its variation is described by an additive upper bound, in which the matrices $\Delta W_b(\theta_b)$ allocate the bound among operators, including cross-slot couplings. The declared family must satisfy the following condition.
\begin{assumption}\label{as:family}
For every applied subset $\mathcal A\subseteq\mathcal B$ and every choice of parameters $\theta_d\in\Theta_d$, the closed-loop and cost matrices satisfy
\begin{gather}
M=M_0+\sum_{d\in\mathcal A}\Delta M_d(\theta_d),\qquad W\preceq W_0+\sum_{d\in\mathcal A}\Delta W_d(\theta_d).
\label{eq:family}
\end{gather}
For each operator $b$, there is a fixed matrix $E_b\succeq0$ such that
\begin{equation}
\Sym(P_0\Delta M_b(\theta_b))+\Delta W_b(\theta_b)\preceq E_b,\qquad \theta_b\in\Theta_b.
\label{eq:defect}
\end{equation}
\end{assumption}

The matrix $E_b$ bounds the contribution of operator $b$ to the certificate inequality. Combining~\eqref{eq:anchor} with Assumption~\ref{as:family} gives
\begin{equation}
\Sym(P_0M)+W\preceq-\Sigma_0+\sum_{b\in\mathcal A}E_b.\label{eq:loss}
\end{equation}
During deployment, $\mathcal A(t)$ and $\theta_d(t)$, hence $M(t)$ and $W(t)$, vary, and Assumption~\ref{as:family} makes~\eqref{eq:loss} valid throughout. The accumulated defects may exhaust the nominal margin, motivating corrections to $P_0$ that accommodate the changes while retaining a prescribed positive margin. Section~\ref{sec:envelope} constructs the bounds in~\eqref{eq:defect}.

\emph{Problem.} Given the inherited pair $(F_0,P_0)$, the declared changes, and the information sets, construct local certificate corrections from data within the read sets, together with a joint admission test that retains a margin $\epsilon I$, where $0<\epsilon\le\lambda_{\min}(\Sigma_0)$. Establish guarantees (i) for a fixed compatible batch of records, one satisfying the joint admission conditions, and (ii) when records are admitted, installed, applied, and retired asynchronously, so that the guarantees cover the mixed configurations reached when operators deploy independently.

\section{Local certificate repair and joint admission}\label{sec:repair}
Solving the problem of Section~\ref{sec:model} requires three ingredients, which we develop in this order: a bound on the effect that each declared change has on the inherited certificate, a local correction of the certificate that absorbs this effect, and a joint admission test that decides which corrections can be used together.
\subsection{Bounding the effect of a local change}\label{sec:envelope}

To make Assumption~\ref{as:family} verifiable, we construct $\Delta M_b$, $\Delta W_b$, and $E_b$ from the declared changes. In a \emph{joint ramp}, each $X\in\{A,Q,N,R,F\}$ equals $X_0+\sum_{b\in\mathcal A}\theta_b\Delta X_b$ with $\theta_b\in[0,1]$. The changes are local: $\Delta A_b$ and $\Delta F_b$ have nonzero blocks only in $\mathcal B_b\times\mathcal D_b$, with $\mathcal D_b$ as in Section~\ref{sec:model}, $\Delta N_b$ is diagonal with support in $\mathcal B_b$, $\Delta R_b$ has square support $\mathcal B_b$, and $\Delta Q_b$ has square support in a declared set $\mathcal Q_b\subseteq\mathcal V$ of users. When $\alpha_F>0$, changes to exposure edges between slots require a bound beyond this recipe. Since $M=A-F$ is linear in the data, the closed loop changes additively, $M=M_0+\sum_{b\in\mathcal A}\theta_b(\Delta A_b-\Delta F_b)$, which gives the first relation in~\eqref{eq:family}. The cost is not additive, because $W(F)$ is quadratic in $F$ and $R_0$ may couple the input costs of different operators. We separate these cross-slot terms by completing the square,
\begin{equation}
\Sym(\Delta F_b^\top R_0\Delta F_d)
\preceq g_{bd}\Delta F_b^\top R_0\Delta F_b
+g_{db}\Delta F_d^\top R_0\Delta F_d,\label{eq:young}
\end{equation}
for any $g_{bd}>0$ with $g_{bd}g_{db}=1$. Set $\kappa_b=1+\sum_{d\ne b:\,(R_0)_{\mathcal B_b\mathcal B_d}\ne0}g_{bd}$. The choice $g_{bd}=1$ needs no agreement between operators and makes $\kappa_b$ one plus the number of operators coupled with $b$ through $R_0$. Collecting terms by order in $\theta_b$ defines the linear, quadratic, and cubic coefficients
\begin{align}
\Lambda_b={}&\Delta Q_b+F_0^\top\Delta R_bF_0
+\Sym(F_0^\top R_0\Delta F_b) \nonumber \\ 
&-\Sym(N_0\Delta F_b+\Delta N_bF_0), \nonumber\\
\Upsilon_b={}&\kappa_b\Delta F_b^\top R_0\Delta F_b
+\Sym((F_0^\top\Delta R_b-\Delta N_b)\Delta F_b), \nonumber\\
\Omega_b={}&\Delta F_b^\top\Delta R_b\Delta F_b.\label{eq:data}
\end{align}
\begin{lemma}\label{lem:envelope}
Under joint ramps, let $\Phi_b=\Lambda_b+(\Upsilon_b)_++(\Omega_b)_+$, with $\Lambda_b$, $\Upsilon_b$ and $\Omega_b$ as in \eqref{eq:data}. Then Assumption~\ref{as:family} holds with $\Theta_b=[0,1]$, $\Delta M_b(\theta_b)=\theta_b(\Delta A_b-\Delta F_b)$, $\Delta W_b(\theta_b)=\theta_b\Phi_b$, and
\begin{equation}
E_b=\big(\Sym(P_0(\Delta A_b-\Delta F_b))+\Phi_b\big)_+.\label{eq:recipe}
\end{equation}
\end{lemma}
\begin{proof}
For $b\ne d$, the supports give $\Delta N_b\Delta F_d=0$ and $\Delta R_b\Delta F_d=0$. Expanding~\eqref{eq:cost} and applying~\eqref{eq:young} to $\theta_b\Delta F_b$ and $\theta_d\Delta F_d$ gives $W-W_0\preceq\sum_{b\in\mathcal A}\big(\theta_b\Lambda_b+\theta_b^2\Upsilon_b+\theta_b^3\Omega_b\big)$.
Since $\theta^jB\preceq\theta B_+$ for symmetric $B$, $\theta\in[0,1]$ and $j=2,3$, this yields $W-W_0\preceq\sum_{b\in\mathcal A}\theta_b\Phi_b$ and $\Sym(P_0\Delta M_b)+\Delta W_b\preceq\theta_bE_b\preceq E_b$.
\end{proof}
The step $\alpha_b$ scales only the gain rows, so $E_b$ must cover the full model and weight change even at $\alpha_b=0$. Appendix~\ref{app:bounds} gives bounds for a model-first transition and for independent model and gain ramps.

\subsection{Finite Gramian repair, support, and data}

By~\eqref{eq:loss}, the change of operator $b$ costs at most $E_b$ of the margin. Replacing $P_0$ by $P_0+X_b$ adds $\Sym(X_bM_0)$ to the left-hand side, plus cross terms $\Sym(X_b\Delta M_d)$, so $X_b$ cancels this loss if $\Sym(X_bM_0)=-E_b$. The solution of this Lyapunov equation, the Gramian $\int_0^\infty e^{M_0^\top t}E_be^{M_0t}\,dt$, is generally dense because $M_0$ couples users along the network. We instead use a finite Smith iteration~\cite{Smith1968} for the Euler-discretized loop defined below. This keeps the correction near the change and leaves an explicit residual for admission.
 
The truncation starts from the users on which $E_b$ is nonzero. A \emph{seed} is a set $\mathcal Z_b\subseteq\mathcal V$ such that $E_b$ vanishes outside $\mathcal Z_b\times\mathcal Z_b$. By~\eqref{eq:recipe}, a valid seed is the union of $\mathcal D_b$, $\mathcal H_b$, $\mathcal Q_b$, the users coupled to $\mathcal B_b$ through $P_0$ or $N_0$, and the feedback sets $\mathcal H_d$ of the operators whose input costs are coupled with those of $b$ through $R_0$.
 
The seed is then propagated backward along the closed loop. Set $\mathcal Z_b^{(-1)}=\varnothing$ and $\mathcal Z_b^{(0)}=\mathcal Z_b$, and obtain $\mathcal Z_b^{(j+1)}$ by adding to $\mathcal Z_b^{(j)}$ every user $i$ with $(M_0)_{li}\ne0$ for some $l\in\mathcal Z_b^{(j)}$, that is, every user whose state directly influences the current set. For a propagation step $h_s>0$, used only in this computation, set $T_0=I+h_sM_0$, an Euler approximation of $e^{h_sM_0}$, and for a depth $k_b\in\mathbb N$ define
\begin{equation}
X_b=h_s\sum_{j=0}^{k_b-1}(T_0^j)^\top E_bT_0^j,\qquad
\mathsf T_b=(T_0^{k_b})^\top E_bT_0^{k_b}.\label{eq:gram}
\end{equation}
Set $\tau_b=\|\mathsf T_b\|$. The \emph{repair support} $\mathcal U_b=\mathcal Z_b^{(k_b-1)}$ contains the users on which $X_b$ acts, and the \emph{read halo} $\mathcal O_b=\mathcal Z_b^{(k_b)}$ contains those on which the residual acts. At depth zero, $X_b=0$ and $\mathsf T_b=E_b$.
\begin{lemma}\label{lem:repair}
$X_b$ and $\mathsf T_b$ are positive semidefinite, with square supports contained in $\mathcal U_b$ and $\mathcal O_b$, and
\begin{equation}
\Sym(X_bM_0)=-E_b+\mathsf T_b-h_sM_0^\top X_bM_0.\label{eq:identity}
\end{equation}
\end{lemma}
\begin{proof}
Each term of~\eqref{eq:gram} is a congruence of $E_b\succeq0$. Factor $E_b=G^\top G$ with $G$ having zero columns outside $\mathcal Z_b$. Since $(T_0)_{li}\ne0$ only if $l=i$ or $(M_0)_{li}\ne0$, right multiplication by $T_0$ adds only the users that influence a nonzero column, so $GT_0^j$ has columns in $\mathcal Z_b^{(j)}$, which gives both supports. Telescoping gives $T_0^\top X_bT_0-X_b=h_s(\mathsf T_b-E_b)$, and expanding $T_0$ gives~\eqref{eq:identity}.
\end{proof}
Thus $X_b$ cancels the defect up to the residual $\mathsf T_b$ and a negative semidefinite term, which only helps. If $h_s$ is small enough that $T_0$ is Schur, $\tau_b$ admits a geometrically decreasing upper bound in $k_b$ while $\|X_b\|$ stays bounded. A deeper repair, however, enlarges $\mathcal U_b$ and can increase the interaction charges below, so increasing $k_b$ does not by itself guarantee admission. The next result lists the data that operator $b$ needs.
\begin{lemma}\label{lem:reads}
(i) In~\eqref{eq:recipe}, $E_b$ is determined by $(P_0)_{\mathcal Z_b\mathcal B_b}$, $(\Delta A_b-\Delta F_b)_{\mathcal B_b\mathcal D_b}$, and $(\Phi_b)_{\mathcal Z_b\mathcal Z_b}$, which uses only the changes of $b$, the rows of $F_0$ of $b$ and of the operators coupled with it through $R_0$, and the columns $\mathcal B_b$ of $R_0$ and $N_0$. (ii) Let $(E_b)_{\mathcal Z_b\mathcal Z_b}=G^\top G$, and set $G_0=G$ and $G_{j+1}=G_j(T_0)_{\mathcal Z_b^{(j)}\mathcal Z_b^{(j+1)}}$. Then $X_b=h_s\sum_{j<k_b}G_j^\top G_j$ and $\mathsf T_b=G_{k_b}^\top G_{k_b}$, where each $G_j^\top G_j$ occupies the rows and columns of $\mathcal Z_b^{(j)}$ and is zero elsewhere, and the recursion uses $M_0$ only on $\mathcal O_b\times\mathcal O_b$. (iii) For every operator $d$, $X_b\Delta M_d$ depends only on $(X_b)_{\mathcal U_b,\mathcal U_b\cap\mathcal B_d}$ and $(\Delta M_d)_{\mathcal U_b\cap\mathcal B_d,:}$, and it vanishes if $\mathcal U_b\cap\mathcal B_d=\varnothing$.
\end{lemma}
\begin{proof}
Claim (i) follows from~\eqref{eq:data}--\eqref{eq:recipe}, since $P_0(\Delta A_b-\Delta F_b)=(P_0)_{:,\mathcal B_b}(\Delta A_b-\Delta F_b)_{\mathcal B_b,:}$ and the seed contains every user coupled to $\mathcal B_b$ through $P_0$. For (ii), $G_j$ is the restriction of $GT_0^j$ to the columns of $\mathcal Z_b^{(j)}$, outside which $GT_0^j$ vanishes by the proof of Lemma~\ref{lem:repair}, and for $j<k_b$ the block $(T_0)_{\mathcal Z_b^{(j)}\mathcal Z_b^{(j+1)}}$ involves only users in $\mathcal Z_b^{(k_b)}=\mathcal O_b$. Claim (iii) follows because $X_b$ has square support $\mathcal U_b$ and $\Delta M_d$ has nonzero rows only in $\mathcal B_d$.
\end{proof}
The repair is therefore computable when these blocks are readable and $\mathcal O_b\subseteq\mathcal S_b$.
 
\subsection{Joint admission and the assembled certificate}
A correction consumes dissipation margin and can interact with the changes of other operators. Corrections must therefore be admitted jointly rather than one at a time. We first define the storage assembled from the installed corrections and then the admission test. The live storage is
\begin{equation}
P=P_0+\sum_{b\in\mathcal J}X_b.\label{eq:storage}
\end{equation}
Since $\mathcal J=\mathcal J(t)$ changes only when corrections are installed or removed, $P=P(t)$ is constant between protocol events. Each correction $X_b$ is built for the nominal loop $M_0$, so every applied change adds the interaction $\Sym(X_b\Delta M_d)$. For each ordered pair, including $d=b$, choose a nonnegative diagonal charge satisfying
\begin{equation}
\Sym(X_b\Delta M_d(\theta_d))\preceq\mathsf C_{db}
\qquad\forall\theta_d\in\Theta_d.\label{eq:cross}
\end{equation}
For example, take $\mathsf C_{db}=\chi_{db}I_{\mathcal U_b\cup\mathcal D_d}$, where $\chi_{db}=\sup_{\theta_d\in\Theta_d}\allowbreak\lambda_{\max}(\Sym(X_b\Delta M_d(\theta_d))_+)<\infty$, and, by Lemma~\ref{lem:reads}(iii), take $\mathsf C_{db}=0$ when $\mathcal U_b\cap\mathcal B_d=\varnothing$. Since $\lambda_{\max}$ is convex, for $\Delta M_d$ affine on a compact polytope the maximum is attained at a vertex. 

Fix a storage cap $\bar\gamma_P\ge\lambda_{\max}(P_0)$. Admission requires the margin ledger and the storage cap
\begin{gather}
\sum_{b\in\mathcal R}\tau_bI_{\mathcal O_b}
+\sum_{b,d\in\mathcal R}\mathsf C_{db}
\preceq\Sigma_0-\epsilon I,\label{eq:ledger}\\
\lambda_{\max}(P_0)+\sum_{b\in\mathcal R:\,i\in\mathcal U_b}\|X_b\|
\le\bar\gamma_P\quad\text{for every user }i.\label{eq:cap}
\end{gather}
Both tests are checked user by user and include admitted records not yet installed, so pending updates cannot claim the same unused margin.
\begin{theorem}\label{thm:assembly}
Under Assumptions~\ref{as:convex}--\ref{as:family}, let the corrections~\eqref{eq:gram} and charges~\eqref{eq:cross} satisfy~\eqref{eq:ledger}--\eqref{eq:cap} for an admitted set $\mathcal R$. Every configuration with $\mathcal A\subseteq\mathcal J\subseteq\mathcal R$ and $\theta_d\in\Theta_d$ satisfies
\begin{equation}
\lambda_{\min}(P_0)I\preceq P\preceq\bar\gamma_PI,\qquad
\Sym(PM)+W\preceq-\epsilon I.\label{eq:cert}
\end{equation}
\end{theorem}
\begin{proof}
Expanding with~\eqref{eq:family}, \eqref{eq:storage}, \eqref{eq:anchor}, \eqref{eq:defect}, \eqref{eq:identity}, and~\eqref{eq:cross}, and discarding $-h_sM_0^\top X_bM_0\preceq0$, gives $\Sym(PM)+W\preceq-\Sigma_0+\sum_{d\in\mathcal A}E_d-\sum_{b\in\mathcal J}E_b+\sum_{b\in\mathcal J}\mathsf T_b+\sum_{b\in\mathcal J,\,d\in\mathcal A}\mathsf C_{db}$. Since $\mathcal A\subseteq\mathcal J$, the defect terms sum to $-\sum_{b\in\mathcal J\setminus\mathcal A}E_b\preceq0$. Since $0\preceq\mathsf T_b\preceq\tau_bI_{\mathcal O_b}$, the remaining nonnegative sums can be extended to $\mathcal R$, and~\eqref{eq:ledger} gives the dissipation inequality. Finally, $0\preceq X_b\preceq\|X_b\|I_{\mathcal U_b}$ and~\eqref{eq:cap} give the storage bounds.
\end{proof}
A configuration is \emph{held} when $\mathcal A$, $\mathcal J$ and every $\theta_d$ stay constant. By~\eqref{eq:cert} and $W\succeq qI$, it is then Hurwitz with $\Pi_F\preceq P$.
\begin{corollary}\label{cor:batch}
Under the assumptions of Theorem~\ref{thm:assembly}, set $P_{\mathcal R}=P_0+\sum_{b\in\mathcal R}X_b$. Along any path with $\mathcal A(t)\subseteq\mathcal R$, $\theta_d(t)\in\Theta_d$, and measurable, locally bounded $M(t)$ and $W(t)$,
\begin{equation}
J_{\mathrm{dep}}:=\int_0^\infty z(t)^\top W(t)z(t)\,dt
\le z(0)^\top P_{\mathcal R}z(0).\label{eq:batchcost}
\end{equation}
The state converges exponentially without a parameter-rate or installation-rate condition.
\end{corollary}
\begin{proof}
Theorem~\ref{thm:assembly} with the hypothetical installed set $\mathcal J=\mathcal R$ gives $\Sym(P_{\mathcal R}M)+W\preceq-\epsilon I$ for every applied subset. Integrating with the constant storage proves~\eqref{eq:batchcost}. Since $W\succeq qI$ and $P_{\mathcal R}\preceq\bar\gamma_PI$, the storage decays at rate at least $(q+\epsilon)/\bar\gamma_P$.
\end{proof}
The common storage is an analysis object, so installation times do not matter for a fixed batch.
\section{Deployment and performance verification}\label{sec:protocol}
Section~\ref{sec:repair} certifies a fixed set of records. We now describe how independent operators admit, install and apply their records so that this certificate is never violated, then show how the guarantee extends when the admitted collection keeps evolving, and finally verify that the final controller reduces the expected cost.
\subsection{Admission, installation, application, and withdrawal}
For each operator $b$, a dependency registry lists every operator $d$ whose records can interact with those of $b$ in either order, based on the largest allowed supports. A conservative choice includes $d$ whenever $\mathcal S_b\cap\mathcal B_d\ne\varnothing$ or $\mathcal S_d\cap\mathcal B_b\ne\varnothing$. Installation publishes a certificate correction but does not change the plant until the corresponding gain or model change is applied.
The shared resources are the record of each operator and the ledger and cap entries of each user, owned by the operator of its slot. A write lock gives an operator the exclusive right to modify a resource, a read lock prevents others from modifying a resource while it is read, and each resource carries a version number that increases at every committed update.
\begin{assumption}\label{as:exec}
(i) The matrices $M(t)$ and $W(t)$ are piecewise continuous and locally bounded in time. The state $z$ is absolutely continuous on bounded intervals, satisfies~\eqref{eq:model} at almost every time, and does not jump at protocol events, of which there are finitely many on every bounded interval. The storage $P(t)$ changes only at these events and takes its new value at the event time. (ii) Record versions, stages, reservations, locks and credits survive crashes. Each owner executes a transition atomically and at most once after checking its record version and stage. The dependency registry lists every operator and user that can interact, and a lock is released only by its holder.
\end{assumption}
Each record proceeds through four rules. (R1)~\emph{Admit:} Operator $b$ write-locks its record, read-locks the records of all dependent operators, and write-locks the ledger and cap entries of the users that may be charged, acquiring all resources in a fixed global order. It rechecks the versions of the data it used, reserves $\tau_bI_{\mathcal O_b}+\mathsf C_{bb}+\sum_{d\in\mathcal R}(\mathsf C_{db}+\mathsf C_{bd})$ on the ledger and $\|X_b\|$ on the cap of each user in $\mathcal U_b$. It admits the record only if~\eqref{eq:ledger}--\eqref{eq:cap} hold and every reservation is acknowledged, and otherwise releases the provisional holds. Locks remain held until commitment or abort and are then released. (R2)~\emph{Install:} When the credit condition of Section~\ref{sec:credits} permits, the operator atomically debits $\log\mu_b$, publishes $X_b$ and adds $b$ to $\mathcal J$. (R3)~\emph{Apply:} After the installation commits, it adds $b$ to $\mathcal A$, and $\theta_b$ may then move anywhere within $\Theta_b$. (R4)~\emph{Withdraw:} After its declared model, weight and gain changes have permanently vanished, the operator removes $b$ from $\mathcal A$ before removing $X_b$ and $b$ from $\mathcal J$. It then retires the record and releases its reservations exactly once under the admission locks. A replacement record requires withdrawal followed by fresh admission.
The protocol preserves the invariant below under message delays, duplications, and losses, but it does not guarantee progress, and by (R4) it does not replace $(F_0,P_0)$ after a permanent change.
\begin{proposition}\label{prop:invariant}
Suppose Assumptions~\ref{as:convex}--\ref{as:family} and item (ii) of Assumption~\ref{as:exec} hold. Starting from empty record sets, rules (R1)--(R4) preserve $\mathcal A\subseteq\mathcal J\subseteq\mathcal R$ and~\eqref{eq:ledger}--\eqref{eq:cap} along every finite sequence of completed transitions.
\end{proposition}
\begin{proof}
The empty configuration satisfies both budgets. Inductively, admission locks fix every interacting record, and both ordered interaction charges are reserved before the new admission becomes visible. Incomplete transactions retain holds without grants, making tests stricter. Installation and application do not change reservations. Withdrawal removes the applied change before its correction and releases only nonnegative charges. Transactions sharing a resource are serialized, while disjoint transactions commute. Durable, version-checked transitions reject duplicate or stale messages and preserve holds after message loss or crashes.
\end{proof}

\subsection{Stability for evolving admitted collections}\label{sec:credits}
Corollary~\ref{cor:batch} applies whenever one compatible batch contains the entire execution. Otherwise the live storage changes as corrections enter and leave. Between events, $V=z^\top Pz$ decreases along trajectories, but an installation can increase it by a factor of at most $\mu_b=1+\|X_b\|/\lambda_{\min}(P_0)$.
Operator $b$ maintains credit $c_b\in[0,\beta_b]$ with refill rate $\varrho_b\ge0$. If $b$ installs nothing in $(s,t]$, then
$c_b(t)=\min\{\beta_b,c_b(s)+\varrho_b(t-s)\}$.
An installation consumes $\log\mu_b$ and waits if the credit is insufficient, so progress requires $\beta_b\ge\log\mu_b$. These credits cap the storage growth caused by the installations of $b$ on any interval $(s,t]$ at the factor $e^{\beta_b+\varrho_b(t-s)}$, and the bound below follows by direct integration between events and summation over installations, as in the average-dwell-time argument of~\cite{Hespanha1999}.
\begin{theorem}\label{thm:main}
Under Assumptions~\ref{as:convex}--\ref{as:exec}, run (R1)--(R4) from empty record sets, with corrections~\eqref{eq:gram}, charges~\eqref{eq:cross}, and initial credits $c_b(0)\in[0,\beta_b]$. Every reached configuration satisfies~\eqref{eq:cert}, and when held it is Hurwitz with $\Pi_F\preceq P$. With $\sigma=(q+\epsilon)/\bar\gamma_P$, $\beta_\Sigma=\sum_b\beta_b$, and $\varrho_\Sigma=\sum_b\varrho_b$, for all $t\ge s\ge0$,
\begin{gather}
V(t)\le e^{\beta_\Sigma}e^{-(\sigma-\varrho_\Sigma)(t-s)}V(s),\label{eq:decay}\\
\|z(t)\|\le\sqrt{\bar\gamma_P/\lambda_{\min}(P_0)}\,
e^{\beta_\Sigma/2}e^{-(\sigma-\varrho_\Sigma)(t-s)/2}\|z(s)\|.\nonumber
\end{gather}
Thus the state converges exponentially if $\varrho_\Sigma<\sigma$.
\end{theorem}
\begin{proof}
Proposition~\ref{prop:invariant} and Theorem~\ref{thm:assembly} give~\eqref{eq:cert}. Since $W\succeq qI$, $\Sym(PM)\preceq-(q+\epsilon)I$, which proves held stability and, by integration, $\Pi_F\preceq P$. Between events, $P$ is constant and $P\preceq\bar\gamma_PI$, so $\dot V=z^\top\Sym(PM)z\le-(q+\epsilon)\|z\|^2\le-\sigma V$. At an installation, $P^-\succeq\lambda_{\min}(P_0)I$ gives $X_b\preceq(\mu_b-1)P^-$, so continuity of $z$ yields $V^+\le\mu_bV^-$. Removals give $P^+\preceq P^-$, and the other events leave $P$ unchanged. Over $(s,t]$, the credit of operator $b$ stays nonnegative, grows at rate at most $\varrho_b$, and drops by $\log\mu_b$ at each installation of $b$, so the sum of these $\log\mu_b$ is at most $c_b(s)+\varrho_b(t-s)\le\beta_b+\varrho_b(t-s)$. Since $(s,t]$ contains finitely many events by Assumption~\ref{as:exec}, integrating $\dot V\le-\sigma V$ between events and summing the logarithms of the jump factors gives $\log V(t)-\log V(s)\le-\sigma(t-s)+\sum_b(\beta_b+\varrho_b(t-s))$ whenever $z(s)\ne0$, which is~\eqref{eq:decay}. The state estimate follows from $\lambda_{\min}(P_0)\|z\|^2\le V\le\bar\gamma_P\|z\|^2$.
\end{proof}
An external change that precedes installation requires a separate certificate. The following result applies only to an initial waiting phase with inherited gains and no installed correction.
\begin{corollary}\label{cor:waiting}
Under Assumptions~\ref{as:convex}, \ref{as:ref} and~\ref{as:exec}(i), let a finite set $\mathcal C$ of operators satisfy~\eqref{eq:family}--\eqref{eq:defect}, with $E_b$ computed for $\Delta F_b=0$ and no correction installed. If, for some $\eta>0$,
\begin{equation}
\sum_{b\in\mathcal C}E_b\preceq\Sigma_0+(q-\eta)I,\label{eq:wait}
\end{equation}
then $\Sym(P_0M)\preceq-\eta I$ for all parameters in the declared ranges, and $z^\top P_0z$ decays at rate at least $\eta/\lambda_{\max}(P_0)$.
\end{corollary}
\begin{proof}
For every applied subset of $\mathcal C$, \eqref{eq:family}--\eqref{eq:defect} and $E_b\succeq0$ give $\Sym(P_0M)+W\preceq-\Sigma_0+\sum_{b\in\mathcal C}E_b$. Subtracting $W\succeq qI$ and applying~\eqref{eq:wait} gives $\Sym(P_0M)\preceq-\eta I$, so $\tfrac{d}{dt}z^\top P_0z\le-\eta\|z\|^2\le-(\eta/\lambda_{\max}(P_0))\,z^\top P_0z$.
\end{proof}
This guarantees stability during the initial waiting phase. For changes within one fixed compatible batch, Corollary~\ref{cor:batch} also supplies a common stability and cost certificate independent of installation times. The online rules still require installation before application.

\subsection{Verifying final expected-cost improvement}\label{sec:performance}
Stability and a bound on the deployment cost do not show that the update is worth deploying. We therefore fix final data $(A_1,Q_1,N_1,R_1)$ and a gain $F_1$ within the admitted ranges. We compare $F_1$ with $F_0$ under these same data, using $M=A_1-F$, $W=W(F;Q_1,N_1,R_1)$ and the corresponding value matrix $\Pi_F$.
\begin{proposition}\label{prop:value}
Let $M=A-F$ and $W=W(F)$, and suppose $\Psi\succ0$ satisfies $-\Sym(\Psi M)\succeq\delta I$ for some $\delta>0$. For any symmetric approximation $Y$ of the value matrix $\Pi_F$, let $B=\Sym(YM)+W$ be its residual, $s=\max_b\sum_d\|B_{\mathcal B_b\mathcal B_d}\|$, $\underline Y=Y-(s/\delta)\Psi$, and $\overline Y=Y+(s/\delta)\Psi$. Then $\underline Y\preceq\Pi_F\preceq\overline Y$. For an initial second moment $\Xi=\mathbb E[z_0z_0^\top]\succeq0$, let $\underline J_F=\tr(\Xi\underline Y)$ and $\overline J_F=\tr(\Xi\overline Y)$. If the bounds for the two gains satisfy $0<\underline J_{F_0}$ and $\overline J_{F_1}<\underline J_{F_0}$, then the final expected cost decreases by at least $1-\overline J_{F_1}/\underline J_{F_0}$ relative to $F_0$.
\end{proposition}
\begin{proof}
Bounding $|x^\top Bx|$ by the symmetric matrix of block norms, whose spectral radius is at most its largest row sum, gives $\|B\|\le s$. Hence $\Sym(\underline Y M)+W\succeq B+sI\succeq0$ and $\Sym(\overline Y M)+W\preceq B-sI\preceq0$. The storage inequality makes $M$ Hurwitz, so integration yields $\underline Y\preceq\Pi_F\preceq\overline Y$. Taking traces against $\Xi\succeq0$ gives the expected-cost bounds and their comparison.
\end{proof}
For $F_1$, admission gives $\Psi=P_{\mathcal R}$ and $\delta=q+\epsilon$. For $F_0$ under $A_1$, the bounds $W_0\succeq q_0I$ and $\|\Sym(P_0(A_1-A_0))\|\le\omega$ give $\Psi=P_0$ and $\delta=q_0+\lambda_{\min}(\Sigma_0)-\omega$, provided $\delta>0$.

Since the proposition holds for any symmetric $Y$, $Y$ can be computed locally. Start from a sparse nominal value approximation computed offline. On selected user pairs, block-Jacobi iteration solves the $(i,j)$ block of $\Sym(YM)+W=0$ for $Y_{ij}=Y_{ji}^\top$, keeping the other blocks at their previous values. The Sylvester equations are uniquely solvable when every $M_{ii}$ is Hurwitz. Each symmetric block pair has one writer, and only pairs within the declared patches are updated. Convergence is unnecessary because the residual bounds the error of the resulting approximation. Residual evaluation uses one frozen symmetric $Y$ and target configuration. An operator reads all blocks entering $B_{\mathcal B_b\mathcal B_d}$ and sends its norm to the block-row owner. Row sums and a network-wide maximum give $s$. For block-diagonal $\Xi$, local trace contributions are summed across operators.

\section{Numerical illustration}\label{sec:example}

We illustrate how much repair is needed to deploy fixed regional LQR updates and how their performance is certified from regional data. Model~\eqref{eq:model}--\eqref{eq:drift} has two topics and $240$ users on a ring with asymmetric weights, partitioned into $60$ slots of four consecutive users. User $i$ is influenced by its neighbors with weights $a_{i,i-1}=.12+.01\,((i-1)\bmod 4)$ and $a_{i,i+1}=.035+.005\,((i-1)\bmod 3)$, with indices modulo $240$. The anchoring strengths are $a_i=.44+.01\,((i-1)\bmod 4)$, and $C=\left[\begin{smallmatrix}.8&.2\\.35&.65\end{smallmatrix}\right]$. We set $Q_0=1.05I$, $N_0=-.45I$, and $R_0=1.15I+.12L_c$, where $L_c$ is the cycle Laplacian on the $480$ state coordinates. We use $g_{bd}=1$, giving $\kappa_b=3$. We obtain the inherited nearest-neighbor gain and block-diagonal storage function, with $2 \times 2$ user-wise blocks, via the dissipative synthesis in~\cite{Mariano2026Structured} with weight $Q_0+.035I$. The certificate~\eqref{eq:anchor} with $\Sigma_0=.035I$ is verified on the full network. We take $\epsilon=.005$, $h_s=.22$, and $\bar\gamma_P=1.15\lambda_{\max}(P_0)\approx.44$.

For each $\ell=0,\ldots,9$, engagement increases by $1.2$ in slots $6\ell+r$, $r\in\{2,3,5,6\}$, lowering their diagonal entries of $N$ by $.6$, while anchoring decreases by $.125$ in slot $6\ell+4$. Assumption~\ref{as:convex} still holds, with $q=1.05-1.05^2/1.15>0$. Each affected operator solves its regional LQR on its four users and one neighbor on either side, so $|\mathcal H_b|=6$. The $50$ proposals use $\alpha_b=.5$ and follow joint ramps with their declared changes. We fix these proposals while varying repair depth, so deeper repair changes only the certificate.

\begin{figure}[!t]
\centering
\includegraphics[width=.85\columnwidth]{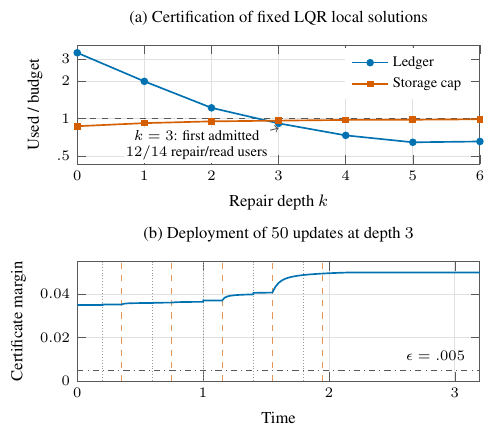}
\caption{Certification and deployment on $240$ users. (a) Fraction of the ledger and storage-cap budgets used at the most loaded user, versus repair depth. Admission requires both fractions to be at most one. (b) Certificate margin at depth $3$. The dotted lines mark the five installation times, the orange dashed lines mark the five ramp starts, each shared by ten slots, and the horizontal line marks the required reserve $\epsilon$.}
\label{fig:performance}
\end{figure}

Using the eight-user support of $E_b$ as seed, depths $0$--$2$ fail admission, whereas depth $3$ passes with $12$ users in each operator's repair support and $14$ in its read halo (Fig.~\ref{fig:performance}(a)). Increasing depth initially reduces ledger use, but enlarges supports and increases storage growth. For the final gain, repair raises the certificate margin $\lambda_{\min}(-\Sym(PM)-W)$ from $-.04$ under $P_0$ to $.05$.

Under the final data with $\Xi=I$, the regional update $F_1$, which changes $400$ of the $480$ gain rows, lowers the held cost from $112.88$ under $F_0$ to $90.22$, an improvement of $20.07\%$. Starting from $Y=P_0$, eight offline sweeps retain value blocks within two user hops. Four online sweeps preserve this mask on six-user patches. With $q_0=1.05-.45^2/1.15$ and $\omega=.25\lambda_{\max}(P_0)$, the outward-rounded bounds $\underline J_{F_0}=112.44$ and $\overline J_{F_1}=98.88$ certify at least $12.06\%$ improvement by Proposition~\ref{prop:value}. Each operator reads at most $14$ users. With eight-user patches, it rises to $18.01\%$ with the same reads but more data exchange. Full-network Lyapunov solves confirm these bounds. A periodic redesign of all $480$ rows with the same feedback allowance costs $80.88$ and has synthesis bound $99.57$, verified on the full network and above the $98.88$ certified for $F_1$.

Within every six-slot group, corrections are installed in slot order $2,3,5,6,4$ at $.20,.60,1.00,1.40,1.80$, so each of these times sees ten simultaneous installations, one per group. Each ramp starts $.15$ after its installation and lasts $.60$, so consecutive ramps overlap. The margin in Fig.~\ref{fig:performance}(b) remains above $.035$, well above the reserve $\epsilon=.005$. Taking expectations, Corollary~\ref{cor:batch} bounds the expected deployment cost by $\tr P_{\mathcal R}\le155.26$, independently of installation times within the declared batch. With $c_b(0)=\beta_b=\log\mu_b$ for each affected slot and no refill, Theorem~\ref{thm:main} gives $V(t)\le e^{1.72}e^{-.22t}V(0)$.

\section{Conclusion}\label{sec:disc}
An inherited quadratic certificate can be adapted through finitely supported positive semidefinite corrections, with residuals and interactions charged against a shared margin. Admission certifies mixed model, objective, and gain configurations. A fixed compatible batch admits a common storage and a deployment-cost bound, while reservations and bounded storage increases ensure stability as the admitted collection evolves. Final expected-cost verification distinguishes a local proposal's benefit from its deployment guarantee. A numerical illustration shows the benefit of the approach on a network of $240$ users undergoing local changes and updates, with a certified cost reduction with respect to the inherited controller of at least $12.06\%$.

The construction is sufficient and requires declared ranges and readable data. Dense inherited certificates or expanding neighborhoods can make repair nonlocal, and greater depth need not yield admission. Tighter interaction charges and adaptive choices of depth and patch size could reduce this conservatism. Further work will tackle these limitations and more: certificate renewal after permanent changes, nonzero anchor opinions that shift the equilibrium, and user populations that change the state dimension. Unexpected changes call for online identification and change detection integrated with admission, leading to a hybrid formulation in which detection triggers protocol events.

\appendices
\section{Separate model and gain ramps}\label{app:bounds}
Lemma~\ref{lem:envelope} assumes that the model and gain changes of a slot progress jointly. Here we let $\nu_b\in[0,1]$ parameterize the changes of $A,Q,N,R$ and $\vartheta_b\in[0,1]$ the change of the gain, so that $\Delta M_b=\nu_b\Delta A_b-\vartheta_b\Delta F_b$. Under the support assumptions of Section~\ref{sec:envelope}, the expansion in the proof of Lemma~\ref{lem:envelope} gives the cost envelope $\Delta W_b={}\nu_b\Lambda_b^{\nu}+\vartheta_b\Lambda_b^{\vartheta}+\vartheta_b^2\kappa_b\Delta F_b^\top R_0\Delta F_b
+\nu_b\vartheta_b\Lambda_b^{\times}+\nu_b\vartheta_b^2\Omega_b$,
where $\Lambda_b^{\nu}=\Delta Q_b-\Sym(\Delta N_bF_0)+F_0^\top\Delta R_bF_0$ and $\Lambda_b^{\vartheta}=\Sym((F_0^\top R_0-N_0)\Delta F_b)$ are the model and gain parts of $\Lambda_b$, and $\Lambda_b^{\times}=\Sym((F_0^\top\Delta R_b-\Delta N_b)\Delta F_b)$, so that $\Upsilon_b=\kappa_b\Delta F_b^\top R_0\Delta F_b+\Lambda_b^{\times}$. With $B_b^{\Delta}=\Sym(P_0\Delta A_b)+\Lambda_b^{\nu}$ and $B_b^{F}=\Lambda_b^{\vartheta}-\Sym(P_0\Delta F_b)$, the total defect is $\nu_bB_b^{\Delta}+\vartheta_bB_b^{F}$ plus the three higher-order terms of $\Delta W_b$.

Three timings are of interest. If both parameters progress jointly, $\nu_b=\vartheta_b=\theta_b$ and~\eqref{eq:recipe} applies. If the model changes first, that is, $\vartheta_b=0$ while $\nu_b$ goes from $0$ to $1$, followed by $\nu_b=1$ while $\vartheta_b$ goes from $0$ to $1$, a valid bound is
\begin{equation*}
E_b=(B_b^{\Delta})_++\big(B_b^{\Delta}+B_b^F+(\Upsilon_b)_++(\Omega_b)_+-(B_b^{\Delta})_+\big)_+.
\end{equation*}
Indeed, the first segment has defect $\nu_bB_b^{\Delta}$, and on the second segment the defect is bounded by $B_b^{\Delta}+\vartheta_b(B_b^F+(\Upsilon_b)_++(\Omega_b)_+)$. The displayed $E_b$ dominates zero and both segment endpoints, hence the entire range. Finally, if the two parameters are independent, each scalar parameter monomial lies in $[0,1]$, and the more conservative bound $ E_b=(B_b^{\Delta})_++(B_b^F)_++\kappa_b\Delta F_b^\top R_0\Delta F_b+(\Lambda_b^{\times})_++(\Omega_b)_+$ is valid. Since $\Delta M_b$ is affine in these parameters, interaction charges can be checked at the vertices of the corresponding rectangle or line segments.

\bibliographystyle{ieeetr}
\bibliography{ecc_references}
\end{document}